\documentstyle[12pt]{article}
\input epsf

\newlength{\extraspace}
\newlength{\extraspaces}
\newcommand{\be}{\begin{equation}
\addtolength{\abovedisplayskip}{\extraspaces}
\addtolength{\belowdisplayskip}{\extraspaces}
\addtolength{\abovedisplayshortskip}{\extraspace}
\addtolength{\belowdisplayshortskip}{\extraspace}}
\newcommand{\ee}{\end{equation}}

\newcommand{\bq}{\begin{eqnarray}
\addtolength{\abovedisplayskip}{\extraspaces}
\addtolength{\belowdisplayskip}{\extraspaces}
\addtolength{\abovedisplayshortskip}{\extraspace}
\addtolength{\belowdisplayshortskip}{\extraspace}}
\newcommand{\eq}{\end{eqnarray}}

\newcommand{\newsection}[1]{
\vspace{15mm}
\pagebreak[3]
\addtocounter{section}{1}
\setcounter{equation}{0}
\setcounter{subsection}{0}
\setcounter{footnote}{0}
\begin{flushleft}
{\large\bf \thesection. #1}
\end{flushleft}
\nopagebreak
\medskip
\nopagebreak}

\begin{document}
\hbox{}
\nopagebreak
\vspace{-3cm}
\addtolength{\baselineskip}{.8mm}
\baselineskip=24pt
\begin{flushright}
{\sc UMN-TH-1317-94}\\
{\sc TPI-MINN-94/37-T}\\
{\sc OUTP}- 94  23 P\\
hep-th@xxx/9410067 \\
 October  1994
\end{flushright}

\begin{center}
{\Large  Compact QED$_3$ - a simple example of a variational calculation
in a gauge theory.}\\
\vspace{0.1in}

{\large Ian I. Kogan}
\footnote{ On  leave of absence
from ITEP,
 B.Cheremyshkinskaya 25,  Moscow, 117259, Russia}\\
{\it  Theoretical
 Physics,
1 Keble Road, Oxford, OX1 3NP, UK}\\
\vspace{0.1in}
and\\
\vspace{0.1in}
{\large Alex Kovner}\\
{\it Physics Department, University of Minnesota\\
116 Church st. S.E., Minneapolis, MN 55455, USA}\\
\vspace{0.1in}

\vspace{0.1in}

{\sc  Abstract}
\end{center}

\noindent
We apply a simple mean field like variational calculation
to compact QED in 2+1 dimensions.
Our variational ansatz explicitly preserves compact
gauge invariance of the theory. We reproduce in this framework
all the known results, including dynamical mass generation,
Polyakov scaling
and the nonzero string tension. It is hoped that
this simple example can be a useful reference point
for applying similar approximation techniques
to nonabelian gauge theories.
\vfill

\newpage
\newsection{Introduction.}

Study of the confining regime in QCD, as well as of many other
strong interaction phenomena requires going beyond the
perturbation theory.
However, application of
analytic nonperturbative methods in quantum field theory,
is a very complicated and not too well developed area. This is
especially true for nonabelian gauge theories.
Recently we have formulated a gauge invariant variational
approach and used a restricted variational ansatz to
study the ground state of a pure  Yang Mills theory in 3+1
dimensions \cite{var}.
As with any new technique, it is desirable to develope an intuition
for it, by first considering simpler examples.

A theory which
posesses many common qualitative features with QCD (such as confinement
and dynamical mass generation), and yet is much simpler and much more tractable
is compact electrodynamics in 2+1 dimensions. Moreover,
this theory has been
previous extensively studied by both, analytical \cite{polyakov}
and numerical \cite{lattice} methods. It seems therefore to be a perfect
testground for application of our variational method.
This is precisely the aim of this note. We will apply a gauge
invariant variational approximation of ref.\cite{var} to this theory.
It is hoped that this toy calculation can teach us something about
improving variational ansatz for realistic 3+1 dimensional nonabelian
theories. It is also a nice excercise in itself, since it
gives a vivid hamiltonian picture of Polyakov's monopole-instanton
condensation phenomenon, which to our knowledge does not exist
in the literature.

The paper is organized as follows. In the rest of this section we discuss
the Hamiltonian formalism for the compact QED$_3$.
In section 2 we set up our variational ansatz and discuss some
important properties of the variational wave functionals.
Section 3 contains calculation of the expectation value of the energy
and solution of minimization equations. We also show that the Wilson loop
in the best variational state has the area law and calculate the string
tension. In section 4 we discuss our results and the interpretation
of our calculation from the point of view of Polyakov's dilute monopole
gas approximation to Euclidean partition function.

The theory is defined by the following Hamiltonian:
\begin{equation}
H=\frac{1}{2}[E_i^2+b^2]
\label{ham}
\end{equation}
The field $b$ is somewhat
different from the usual magnetic field $B=\epsilon_{ij}\partial_iA_j$.
We will explain its definition in
a short while.
All the physical states should satisfy the Gauss'
law constraint
\begin{equation}
\exp\{i\int d^2x\partial_i\lambda(x)E_i(x)\}|\Psi>=|\Psi>
\label{constr}
\end{equation}
One should note that there is a crucial difference between the Gauss' law
in the compact theory and in the noncompact one.
In the noncompact theory equation (\ref{constr}) should be satisfied only
for regular functions $\lambda$.
For example, the operator
\begin{equation}
V(x)=\exp\left\{\frac{i}{g}\int d^2y\frac{\epsilon_{ij}(x-y)_j}{(x-y)^2}
E_i(y)\right\}
\label{vortex}
\end{equation}
which has the form of  (\ref{constr}) with the function $\lambda$
proportional to the planar angle $\theta$, $\lambda=\frac{1}{g}
\theta(x)$, does not act trivially on physical states.
In fact, this operator creates pointlike magnetic vortices with magnetic flux
$2\pi/g$ \cite{vort} and therefore changes the physical state on which
it is acting.

In the compact theory the situation in this respect is quite
different. Pointlike
vortices with quntized magnetic flux $2\pi n/g$ can not be detected
by any measurement. In Euclidean
path integral formalism of ref.\cite{polyakov} this is the statement
that the Dirac string of the monopole is unobservable and does not cost
any (Euclidean) energy. In the Hamiltonian formalism this translates
into the requirement that
the creation operator of a pointlike vortex
must be indistinguishible from
the unit operator. In other words, the operator (\ref{vortex})
generates a transformation which belongs to the compact gauge group,
and should therefore
act trivially on all physical states. Equation (\ref{constr})
 should therefore
be satisfied also for these operators.

Accordingly, the Hamiltonian of the compact theory also must be invariant
under these transformation. The magnetic field defined as
$B=\epsilon_{ij}\partial_iA_j$, on the other hand
does  not commute with $V(x)$,
\begin{equation}
V^\dagger (x)B(y)V(x)=B(y)+\frac{2\pi}{g}\delta^2(x-y)
\label{commut}
\end{equation}
The Hamiltonian
should therefore
contain not $B^2$ but rather only its singlet part. This is the meaning
of the field $b$ in equation (\ref{ham}). Formally
\begin{equation}
b^2=PB^2P
\label{b}
\end{equation}
where $P$ is the projection operator on the whole compact gauge group,
which includes $V(x)$.
This form is convenient for the purposes of the present calculation, and
we therefore do not write down a more explicit expression for
$b^2$\footnote{In the standard Euclidean lattice formulation
of the theory the potential energy has the form $\cos(ga^2B)$, where $a$
is the lattice spacing. This obviously has the same property as $b^2$,
that is invariant under the
transformation eq. \ref{commut} and reduces to it in the weak
coupling limit. Our definition eq. \ref{b} is equivalent to the Villain
form of the action.}.

\newsection{The variational ansatz.}
Our aim in this paper is to find a vacuum wave functional of this
theory. Following Polyakov, we will be working in the weakly coupled regime.
Since the coupling constant $g^2$ in 2+1 dimensions has dimension of mass,
weak coupling means that the following dimensionless ratio is small
\begin{equation}
\frac{g^2}{\Lambda}<<1
\end{equation}
Here $\Lambda$ is the ultraviolet cutoff in the momentum space, which as
always has to be introduced to regularize a quantum field
theory\footnote{If one considers  compact $U(1)$ as an unbroken sector of the
spontaneously broken $SU(2)$  (Georgi-Glashow model), the ultraviolett
cutoff $\Lambda$ will be proportional to the scale of the symmetry breaking.
To be more precise it is  by  order of magnitude  of the  charged vector
boson mass $M_{W}$.}.
For a weakly coupled theory one expects the vacuum wave functional (VWF)
to be not too different from the vacuum of a free theory. Since the  VWF
of free (noncompact) electrodynamics is gaussian in the field basis,
\begin{equation}
\psi[A_i]=\exp\left\{-1/2\int d^2xd^2y A_i(x)G^{-1}_{ij}(x-y)A_j(y)\right\}
\end{equation}
Gaussian variational approach in this case should give a good approximation.
An important caveat, however is that the ground state WF should be gauge
invariant under the full compact gauge group. As a result
it turns out that one can not take just a
Gaussian in $A_i$, since this will not
preserve gauge invariance. The simplest generalization of the Gaussian ansatz
which we use along the lines of \cite{var}, is to project a gaussian WF
into the gauge invariant subspace of the Hilbert space.
We therefore take as our variational anzatz the following set of states
\begin{equation}
\Psi[A_i]=\int D\phi
\exp\left\{-\frac{1}{2}\int d^2xd^2y \left[A_i(x)-\frac{1}{g}\partial_i
\phi(x)\right]G^{-1}(x-y)\left[A_i(y)-\frac{1}{g}\partial_i\phi_i(y)\right]
\right\}
\label{wf}
\end{equation}
The functional integral is over the phase function $\phi(x)$, and
correspondingly the derivatives of $\phi$ in the exponential are understood
modulo $2\pi$. That is, these derivatives do not feel quantized discontinuities
in $\phi(x)$.
The mathematically more precise way to write this
is to  substitute $\partial_i\phi(x)$ by
$\exp\{-i\phi(x)\}\partial_i\exp\{i\phi(x)\}$.
 We will use however, the above
shorthand notation for convenience.
The simple rotational structure of $G_{ij}=\delta_{ij}G$ that appears in the
variational wave functional (\ref{wf})
 is consistent with perturbation theory, as
discussed in \cite{var}.

The ansatz  (\ref{wf}) depends on one function $G(x)$.
We now have to calculate the expectation value of the energy
 in this state, and then minimize it with respect to $G(x)$.

Before proceeding with the calculation we make the following (obvious)
comment. The trial wave functional (\ref{wf})
has a simple interpretation from
the point of view of states of a noncompact theory.
To see this  let us rewrite the functional measure $D\phi$ in a slightly
different way. Any angular function $\phi(x)$ can be parametrized as
\begin{equation}
\phi(x)=\tilde\phi(x)+\phi_v(x)
\end{equation}
where $\tilde\phi(x)$ is a smooth function and $\phi_v(x)$ contains all the
discontinuities and can be written as
\begin{equation}
\phi_v(x)=\sum_{\alpha=1}^{n_+}\theta(x-x_\alpha)-\sum_{\beta=1}^{n_-}
\theta(x-x_\beta)
\label{phiv}
\end{equation}
 and $\theta(x-x_\alpha)$ is a polar angle on plane with
 a center at  $x_\alpha$.  The functional measure  can be written as
\begin{eqnarray}
\int D\phi&=&\int D\tilde\phi\sum_{n_+=1}^{\infty}\sum_{n_-=1}^{\infty}
\sum_{\{x_\alpha\}}\sum_{\{x_\beta\}}\\ \nonumber
&=&
\int D\tilde\phi\sum_{n_+=0}^{\infty}\sum_{n_-=0}^{\infty}\frac{1}{n_+!n_-!}
\prod_{\alpha =1}^{n_{+}}\prod_{\beta=1}^{n_{-}}
\left(\int d^2x_\alpha d^2x_\beta \Lambda^4\right)
\label{measure}
\end{eqnarray}
where in the last equality we have just substituted
integration over the coordinates of vortices and antivortices for
summation, and for this reason introduced explicit  UV cutoff $\Lambda$.
We also introduced term with $n_{+} = n_{-} = 0$ corresponding to the
 absence of vortices.

Let us define the function
\begin{equation}
\tilde\Psi[A]=\int D\tilde\phi\exp\left\{-\frac{1}{2}\int d^2xd^2y
\left[A_i(x)-\frac{1}{g}
\partial_i
\tilde\phi(x)\right]G^{-1}(x-y)\left[A_i(y)-\frac{1}{g}\partial_i
\tilde\phi_i(y)\right]\right\}
\end{equation}
which differs from $\Psi$ in that the integration is
performed only over continuous gauge functions $\tilde\phi$. Obviously,
$\tilde\Psi$ is invariant under noncompact gauge group, and therefore belongs
to the Hilbert space of the noncompact theory. When acting on it,
the vortex operator $V(x)$
defined in (\ref{vortex})
just shifts $A_i(y)$ by $\frac{1}{g}\partial_i\theta(x-y)$.
We have therefore the following representation for $\Psi[A]$
\begin{equation}
\Psi[A]=\sum_{n_{+},n_{-}=0}^\infty\prod_{\alpha = 1}^{n_{+}}
\prod_{\beta= 1}^{n_{-}}V(x_\alpha)
V^{*}(x_\beta)\tilde\Psi[A]
\end{equation}
This representation makes explicit the fact that a WF of the compact theory
is constructed from a WF of the noncompact theory by taking a superposition
of arbitrary number of vortices and anti vortices at every point. This
superposition is obviously invariant under multiplication by a vortex operator
and therefore is its eigenfunction with eigenvalue 1.

Having noted this, we now proceed to calculation of expectation values in
the trial state (\ref{wf}) and the minimization of the
vacuum expectation value of energy.

\newsection{The energy minimization.}

Let us for convenience introduce the following notation
\begin{equation}
A_i^\phi(x)= A_i(x)-\frac{1}{g}\partial_i\phi(x)
\end{equation}
We will also switch to the matrix notations in the following, so that
\begin{equation}
A_iMA_i=\int d^2xd^2yA_i(x)M(x-y)A_i(y)
\end{equation}
The expectation value of any operator in the WF (\ref{wf}) is calculated as
\begin{equation}
<O>=Z^{-1}\int D\phi' D\phi'' DA_i
\exp\left\{-\frac{1}{2} A_i^{\phi'}G^{-1}A_i^{\phi'}\right\}
O(A)
\exp\left\{-\frac{1}{2} A_i^{\phi''}G^{-1}A_i^{\phi''}
\right\}
\label{exp1}
\end{equation}
Here $Z$ is the normalization factor, which is just the norm of the
trial wave functional (\ref{wf}).
Further, if the operator $O$ is explicitly gauge invariant, we may shift the
integration variable $A_i\rightarrow A_i^{\phi''}$, and
reduce this expression to
\begin{equation}
<O>=Z^{-1}\int D\phi D\eta DA_i
\exp\left\{-\frac{1}{2} A_i^\phi
G^{-1}A_i^\phi\right\}O(A)
\exp\left\{-\frac{1}{2} A_iG^{-1}A_i\right\}
\label{exp}
\end{equation}
where we have defined $\phi=\phi'-\phi''$ and $\eta=\phi'+\phi''$. The
integral over $\eta$ is trivial and just gives the volume of the gauge group.
Since the same integral exactly enters $Z$, it always cancells between the
numerator and denominator, and we shall omit it in the following.

As a first step let us calculate the normalization factor $Z$.
\begin{equation}
Z=\int D\phi DA_i
\exp\left\{-\frac{1}{2}\left[ A_i^{\phi}G^{-1}A_i^{\phi}+
A_iG^{-1}A_i\right]\right\}
\label{z}
\end{equation}
The integral over $A_i$ is gaussian and can be trivially performed.
The integral over the noncompact part of the gauge group $\tilde \phi$
is also gaussian.
Then
\begin{equation}
Z=Z_aZ_\phi Z_v
\end{equation}
with
\begin{eqnarray}
Z_a&=&\det (\pi G)\\ \nonumber
Z_\phi&=&\int D\tilde\phi ~\exp\left\{
-\frac{1}{4g^2}\int \partial_i\tilde\phi
G^{-1}\partial_i\tilde\phi\right\}=
\det\left[4\pi g^{2}\frac{1}{\partial^2}G\right]^{1/2}\\
\nonumber
Z_v&=&\int D\phi_v\exp\left\{-\frac{1}{4g^2}\int \partial_i\phi_v
G^{-1}\partial_i\phi_v\right\}
\end{eqnarray}

Using equations (\ref{phiv}) and (\ref{measure}), one can represent $Z_v$
as a partition function of the gas of vortices
\begin{eqnarray}
Z_v&=&\sum_{n_+,n_- = 0}^{\infty}
\prod_{\alpha=1}^{n_{+}}\prod_{\beta=1}^{n_{-}}
dx_\alpha dx_\beta z^{n_++n_-} \\
&\exp&\left\{-\frac{1}{4g^2}[\sum_{\alpha,\alpha'}D(x_\alpha-x_{\alpha'})
+\sum_{\beta,\beta'}D(x_\beta-x_{\beta'})-
\sum_{\alpha,\beta}D(x_\alpha-x_{\beta})]\right\} \nonumber
\label{part}
\end{eqnarray}
The vortex - vortex interaction potential $D(x)$ and the vortex
fugacity $z$ are given by
\begin{eqnarray}
D(x)&=&8\pi^2\int \frac{d^{2}k}{(2\pi)^{2}}\frac{1}{k^2}G^{-1}(k)
\cos(kx)\\ \nonumber
z&=&\Lambda^2\exp\{-\frac{1}{8g^2}D(0)\}
\label{potential}
\end{eqnarray}
Here $G(k)$ is the Fourier transform of the variational ``propagator''
$G(x)$.

The formula
(\ref{part}) reminds one of Polyakov's partition
function of the monopole gas \cite{polyakov}. One should keep
in mind, however that the physical meaning of it is quite different.
The gas described by equation  (\ref{part}) is two dimensional and not
three dimensional, and the interaction between the particles
is not Coulomb, but rather depends on the variational function $G$.
Nevertheless, it has a definite relation with Polyakov's gas
of monopoles and we will discuss this point in the last section
of the paper.

Since $D(0)$ is singular, the last equation should be understood,
as usual in the regularized sence, that is at finite UV cutoff,
$D(0)$ should be substituted by $D(x=1/\Lambda)$.
Note that the variational function $G$ explicitly appears in the
vortex - vortex potential.
Since we expect the UV behavior of $G(k)$ to be the same as in the
free theory $(G^{-1}(k)\rightarrow k)$, we have
\begin{equation}
z=\Lambda^2\exp\{-\frac{\pi}{2}\frac{\Lambda}{g^2}\}
\label{zpar}
\end{equation}

In the following we will need to calculate correlation functions of the
vortex density. To facilitate this we use the standard trick \cite{polyakov},
\cite{samuel} to rewrite the partition function $Z_v$
in terms of a path integral over a scalar field $\chi$.
Let us introduce the vortex density $\rho(x)$
\begin{equation}
\rho(x)=\sum_{\alpha,\beta}\delta(x-x_\alpha)-\delta(x-x_\beta)
\end{equation}
The exponential factor in eq. \ref{part} (including the factors of fugacity)
 can then be rewritten as
\begin{equation}
\Lambda^{2(n_++n_-)}\int D\chi\exp\{-2g^2\chi D^{-1}\chi+i\rho\chi\}
\end{equation}
The summation over the number of vortices is trivial and gives
\begin{equation}
Z_v=\int D\chi\exp\{-2g^2\chi D^{-1} \chi+
\int _x2\Lambda^2\cos\chi(x)\}
\label{sine}
\end{equation}
To calculate the correlator of $\rho$ one can add $i\rho J$ to the vortex
free energy, and calculate functional derivatives of the resulting partition
function with respect to $J$ at zero $J$. A simple derivation gives
\begin{equation}
<\rho(x)\rho(y)>=4g^2D^{-1}(x-y)-16g^4<D^{-1}\chi(x)D^{-1}\chi(y)>
\end{equation}

The propagator of $\chi$ is easily calculated. At weak coupling
$z$ is very small, and all our calculations will be performed
to first order in $z$.
To this order the only contribution comes from the tadpole diagrams. This
is easily seen by rewriting the cosine potential
in equation (\ref{sine}) in the normal ordered form
\footnote{The normal ordering is
performed relative to the free
theory defined by the quadratic part of the action in equation (\ref{sine}).}
\begin{equation}
\cos \chi=<\cos \chi>_0:\cos \chi:=\frac{z}{\Lambda^2}:\cos \chi:
\label{normalorder}
\end{equation}
Therefore to first order in $z$
\begin{equation}
\int d^2 x e^{ikx}<\chi(x)\chi(0)>=
\frac{1}{4g^2D^{-1}(k)+2z}=\frac{D(k)}{4g^2}
-z\frac{D^2(k)}{8g^4} + o(z^2)
\end{equation}
The correlator of the vortex densities is then
\begin{equation}
K(k)=\int d^2x e^{ikx}<\rho(x)  \rho(0)>=2z + o(z^2)
\label{rhoc}
\end{equation}
and in this approximation does not depend on momentum, the $k$-dependence
 will appear in  $z^{2}$ and higher order terms.

 Now we are ready to calculate
the expectation value of the Hamiltonian (\ref{ham}).
First, consider the electric part
\begin{eqnarray}
<\int d^2x E_i^2> = -<\int d^2x \frac{\delta^2}{\delta A_i^2}> ~~~~~~~~~~~~
{}~~~~~~~~~~~~~
\nonumber \\
=  Z^{-1}\int D\phi DA_i
\exp\left\{-\frac{1}{2}A_i^{\phi}G^{-1}A_i^{\phi}\right\}
\left[2{\rm tr} G^{-1}-A_iG^{-2}A_i\right]
\exp\left\{-\frac{1}{2} A_i G^{-1}A_i\right\} ~~~~~~~~
  \\
 = {\rm tr} G^{-1}-\frac{1}{4g^2}Z_\phi^{-1}Z_v^{-1}\int D\phi ~
  \left(\partial_i\phi G^{-2}\partial_i\phi\right)  \exp
\left\{-\frac{1}{4g^2}\partial_i\phi G^{-1}\partial_i\phi\right\}
{}~~~~~~~~~~~~~~~~~~~~~~~~
\nonumber
\end{eqnarray}
Performing the gaussian integral over $\tilde\phi$, this reduces to
\begin{eqnarray}
V^{-1}<\int E^2_i>&=&\frac{1}{2}\int \frac{d^2k}{(2\pi)^{2}}
G^{-1}(k)-\frac{\pi^2}{g^2}
\int \frac{d^2k}{(2\pi)^{2}}  k^{-2}G^{-2}(k)K(k)\\ \nonumber
&=&\frac{1}{2}\int \frac{d^2k}{(2\pi)^{2}}  G^{-1}(k)-\frac{2\pi^2}{g^2}
z\int \frac{d^2k}{(2\pi)^{2}}  k^{-2}G^{-2}(k)
\label{elen}
\end{eqnarray}
Let us note that it is only because of  vortices one has negative
contribution to the electric part of the energy.

Now for the magnetic part.
Since $b^2$ is the singlet part of $B^2$, by definition in every gauge
invariant state $<b^2>=<B^2>$. We will therefore calculate $<B^2>$.
Here one should be a little careful. Since $B^2$ itself is not
gauge invariant, one can not use equation (\ref{exp}),
but rather explicitly keep
both integrals, over $\phi$ and $\eta$.
\begin{eqnarray}
<b^2>&=&Z^{-1}\int D\phi' D\phi'' DA_i
B^2 \exp\left\{-\frac{1}{2}\left[A_i^{\phi'}G^{-1}A_i^{\phi'}+
A_i^{\phi''}G^{-1}A_i^{\phi''}\right]\right\}
\\ \nonumber
&=&Z^{-1}\int D\phi D\eta DA_i
\left[\epsilon_{ij}\partial_i\left(A_j-\frac{1}{2g}(\phi-\eta)\right)
\right]^2
\exp\left\{-\frac{1}{2} \left[A_i^{\phi}G^{-1}A_i^{\phi}+
A_iG^{-1}A_i \right]\right\}
\end{eqnarray}
where, the factor $Z$ contains an extra factor of the volume of the
gauge group relative to equation (\ref{z}), and
as previously $\phi=\phi'-\phi''$ and $\eta=\phi'+\phi''$.
The linear term in $\eta$ vanishes due to the symmetry of the
measure under transformation $\eta\rightarrow -\eta$.
The term quadratic in $\eta$ is independent of the variational parameter $G$.
It does not contribute to the minimization equations, and we omit it in the
following.
We therefore obtain
\begin{eqnarray}
<b^2>&=&Z^{-1}\int D\phi D\eta DA_i
\left[\epsilon_{ij}\partial_i\left(A_j-\frac{1}{2g}\phi\right)\right]^2
\exp\left\{-\frac{1}{2} \left[A_i^{\phi}G^{-1}A_i^{\phi}
+A_iG^{-1}A_i\right]\right\}
 \nonumber \\
&=&Z_a^{-1}\int DA_i [\epsilon_{ij}\partial_iA_j]^2\exp\{-A_iG^{-1}A_i\}=
\frac{1}{2}\int d^2k k^2G(k)
\end{eqnarray}
This is the same result as in the noncompact theory. In fact, this is precisely
what one expects, since a compact state $\Psi$ differs from a noncompact one,
$\tilde \Psi$ only
by the presence of vortices, but $b^2$ by definition should not feel their
presence.

Summarizing, we have the following expression for the expectation value
of the energy density
\begin{equation}
\frac{1}{V}<H>=
\frac{1}{4} \int\frac{ d^2k}{(2\pi)^2}
\left[G^{-1}(k)+k^2G(k) -\frac{4\pi^2}{g^2}
zk^{-2}G^{-2}(k) \right]
\end{equation}
The minimization equation is
\begin{equation}
\frac{1}{4}\left[k^2-G^{-2}(k)\right]+\frac{\pi^2}{g^2}
 \left[ 2zk^{-2}G^{-3}(k)-4\pi^2
\frac{\delta z}
{\delta G(k)}
\int \frac{d^2p}{(2\pi)^2} p^{-2}G^{-2}(p) \right]=0
\label{minim}
\end{equation}

{}From equation (\ref{potential}) one finds
\begin{equation}
\frac{\delta z}{\delta G(k)}=\frac{1}{4g^2}k^{-2}G^{-2}(k)z
\end{equation}

Assuming perturbative behaviour of $G$ at large momenta $(G(k)
\rightarrow k^{-1})$,
the ratio of the fourth term in equation (\ref{minim}) to the third term
is of order
\begin{equation}
\frac{\delta z}
{\delta G(k)}
\frac{\int d^2p p^{-2}G^{-2}(p)}{2zk^{-2}G^{-3}(k)}
\propto\frac{\Lambda^2}{g^2k}
\end{equation}

At weak coupling this is
much greater than one for any value of momentum. We can therefore
omit the third term from equation (\ref{minim}) and get a very
 simple equation on $G^{-2}(k)$
\begin{equation}
k^2-G^{-2}(k) =
  \frac{4\pi^4}{g^{4}} z   k^{-2} G^{-2}(k)
\int \frac{d^2p}{(2\pi)^2} p^{-2}G^{-2}(p)
\end{equation}
 with the  solution
\begin{equation}
G^{-2}(k)=\frac{k^4}{k^2+m^2}
\label{solution}
\end{equation}
where
\begin{equation}
m^2=\frac{4\pi^4}{g^4}z\int \frac{d^2k}{(2\pi)^2}
k^{-2}G^{-2}(k)
\label{mass}
\end{equation}
Using equations (\ref{potential}) and (\ref{solution}) we  get
an equation on mass
\begin{equation}
m^2=\frac{4\pi^4}{g^4} \Lambda^{2}
\exp\left(-\frac{\pi^{2}}{g^{2}}
\int \frac{d^2p}{(2\pi)^2} \frac{1}{\sqrt{p^2 + m^{2}}}\right)
\int \frac{d^2k}{(2\pi)^2}
\frac{k^2}{k^2+m^2}
\label{complicatedm}
\end{equation}
which in case of weak coupling $\Lambda/g^{2} >> 1$ can be simplified
as
\begin{equation}
m^2 = \pi^3 \frac{\Lambda^4}
{g^4}\exp\{-\frac{\pi}{2}\frac{\Lambda}{g^2}\}
\end{equation}
and because $m << g^{2} << \Lambda$ one indeed could neglect the $m$-dependence
in the right hand side of equation (\ref{complicatedm}).
It is clear, that $m$ is precisely the mass gap of the theory. Calculating,
for example the propagator of magnetic field, we find
\begin{equation}
\int d^2xe^{ikx}<b(x)b(0)>= \frac{1}{2}(k^2+m^2)^{1/2}
\end{equation}

Note, that the dynamically generated mass we obtain in
our approximation agrees with Polyakov's result \cite{polyakov}.

Perhaps the most interesting question is whether our best variational VWF
is confining. To answer this question we calculate the expectation
value of the Wilson loop.
\begin{equation}
W_C= <\exp\{ilg\oint_C A_{i}dx_{i}\}> = <\exp\{ilg\int_SBdS\}>
\end{equation}
where $l$ is an arbitrary integer and the integral is over the area $S$
bounded by the loop $C$.
We have written $B$ rather than $b$, since this exponential operator is
invariant under transformations $B(x)\rightarrow B(x)+\frac{2\pi}{g}$,
generated by the vortex operator.
\begin{eqnarray}
W_C=&Z_a^{-1}&\int DA_i
\exp\left\{
-A_iG^{-1}A_i+ilg\int_S\epsilon_{ij}\partial_iA_jd^2x\right\}\\ \nonumber
&Z_\phi^{-1}Z_v^{-1}&\int D\phi\exp\left\{-\frac{1}{4g^2}\partial_i\phi
G^{-1}\partial_i\phi+i\frac{l}{2}\oint_C\partial_i\phi dx_i\right\}
\label{wilson}
\end{eqnarray}

The first factor in weak coupling is simply
\begin{equation}
W_0=\exp\left\{-\frac{l^2}{2}g^2\int_{x,y}<B(x)B(y)>d^2x d^2y\right\}
\label{wa}
\end{equation}
where the integral over both $x$ and $y$ is over the area $S$.
In the limit of large $S$ the leading piece in the exponential
is
\begin{equation}
\frac{l^2}{2}g^2S\lim_{k\rightarrow 0}\frac{k^2}{2}G(k)=\frac{l^2}{4}g^2mS
\end{equation}
This term therefore leads to the area law behaviour and
gives the string tension
\begin{equation}
\sigma=\frac{l^2}{4}g^2m
\label{str}
\end{equation}
and we see that area law $W_{0} \sim \exp(-\sigma S)$ is a direct
 consequence of the nonzero mass gap $m$.

The second factor in equation (\ref{wilson}) is different from unity only
for odd $l$, since $\oint\partial_i\phi dl_i=2\pi(n_+-n_-)$,
where $n_+$ ($n_-$) is the number of vortices (antivortices) inside the loop.
For odd $l$ it can be easily calculated
\begin{equation}
W_v=<\exp\{i\pi\int_S\rho(x)d^2x\}>=\int D\chi \exp\{-2g^2
\chi D^{-1}\chi+\int _{x}2
\Lambda^2\cos(\chi(x)-\alpha(x))\}
\label{wvort}
\end{equation}
where $\alpha(x)$ is the function, which vanishes
for $x$ outside the loop, and is equal to $\pi$ for $x$ inside the loop.
At small coupling this can be
calculated in the steepest descent approximation.
The solution to the classical equations which contributes to the leading order
result is $\chi(x)=0$. For this solution
\begin{equation}
W_v=\exp\{-4zS\}
\label{wvortf}
\end{equation}
where we again used  the normal ordering prescription as in
 (\ref{normalorder}). Clearly this is a subleading correction to
 string tension (\ref{str}), since $z\propto m^2$ and
 $z/g^{2}m \sim \sqrt{z/\Lambda^{2}} \sim \exp(-\pi\Lambda/4g^{2}) <<1$.
With this exponential accuracy, the string tension is therefore given
in our approximation by equation (\ref{str}).

\newsection{Discussion.}

We have presented a simple variational calculation in the compact QED$_3$.
Our variational ansatz was the direct adaptation of the ansatz of
ref.\cite{var} to this theory. The trial wave functionals are explicitly
gauge invariant under the compact gauge group. The integration over the
gauge group is directly responsible for nontrivial dependence of the
energy expectation value on the variational parameters, which leads
to the generation of scale in the best variational state.
The correlators and the Wilson loop calculated in the best variational
state agree with known results.

It is illuminating at this point to interpret our calculation from the
point of view of the three dimensional Euclidean path integral.
The vacuum wave functional of the theory can be represented in path
integral formalism. To get the vacuum WF $\Psi[A]$ one should calculate the
path integral over the fields $A(x,t)$, with $t$ varying from
$-\infty$ to $0$, with the boundary condition $A(x, t=0)=A(x)$.
To be more precise, in calculating VEV of some operator $O(t=0)$,
one should split the time coordinate of the
plane with the time coordinate of the operator, so that one considers
$\Psi[A(x,t=-\epsilon)]$ and $\Psi^*[A(x,t=\epsilon)]$ in the limit
$\epsilon\rightarrow 0$.

The basic objects that appear in the Euclidean path integrals are
monopoles, which in 3D are not propagating particles, but rather
instantons. When described in terms of the vector potential,
or noncompact field strength, a monopole has a Dirac string attached to it.
It is clear that the vortices (antivortices)
of the gauge function
$\phi(x)$, in equation (\ref{wf}) correspond precisely to the intersections
of the Dirac strings of the 3D monopoles (antimonopoles)
with the equal time plane at $t=0$.
The positions of the Dirac strings
are not physical in the compact theory, and only the position of the
monopole itself is gauge invariant.
In fact, for all monopoles that do not sit in the infinitesimally thin
time slab between the planes
$t=-\epsilon$ and $t=\epsilon$, one can always choose the
direction of the Dirac string such that it does not intersect the two
planes. This precisely corresponds to expression equation (\ref{exp}).
 The  combination that enters this path integral nontrivially is
 $\phi=\phi'-\phi''$. At the points,
where both functions $\phi'(x)$ (which corresponds to $\Psi$) and
$\phi''(x)$ (which corresponds to $\Psi^*$) have a vortex, $\phi(x)$ is
regular. This is the situation, when a Dirac string intersects
both planes $t=\pm\epsilon$. When a 3D monopole sits in the slab, only
one of the functions $\phi'$ or $\phi''$ has nonzero vorticity, and so does
$\phi$. The integration over $\phi(x)$ in equation (\ref{exp})
 can be interpreted
therefore as the direct contribution to the expectation value due to the
monopoles  at precisely the time $t=0$.

The fact that in this way one sees directly only the monopoles at
$t=0$, does not mean of course that other monopoles are not taken
into account in this approximation. Indeed, the ``bare'' interaction
potential between the $t=0$ monopoles is $D(x)$ of equation
 (\ref{potential}).
In the best variational state it is already short range, as follows from
the solution for $G(k)$ (see equation (\ref{solution})).
This is in accordance with the 3D picture, that the 3D monopole gas
produces screening. Obviously, if one only looks at the thin slab,
every monopole there will have an antimonopole partner, which sits
nearby (inside the screening length) in the third direction. The 2D monopole
gas will therefore be screened by the 3D interaction, even before the
interaction of the 2D monopoles between themselves
is taken into account. This is perfectly consistent with our
calculation. It is
interesting to note, that even though this 2D interaction
produces additional screening (the cosine term in the effective
theory  (\ref{sine})),
 it is the 3D screening that is responsible for the
area law of the Wilson loop, as is clear from the calculation
of the string tension in  (\ref{wilson}).
In fact, if one takes for $G(k)$ the noncompact expression,
both,
the leading part $W_0$, equation (\ref{wa}), and the
subleading part $W_v$,  (\ref{wvort})
have the perimeter law behaviour. The string tension in $W_0$
vanishes, because in this case
$\lim_{k\rightarrow 0} k^2 G(k) = 0$. In $W_v$
in this case $D^{-1}(k^2=0)=0$, and for large loops the classical
equations of motion, which follow from
equation  (\ref{wvort}), apart from $\chi=0$ have
another solution, which leads to the perimeter dependence of $W_v$.
The existence of the cosine term in the interaction (which appears due to
the 2D screening) does not preclude the existence of this extra solution.

Finally, we note that the calculation presented here can be extended
to compact QED in 3+1 dimensions. In this case the vortex gas part of the
partition function will be replaced by the gas of vortex loops with
the interaction between the loop elements dependent on the variational
function $G$. It is well known that compact $QED_4$ has a phase transition
at a finite coupling constant. There is a good chance that this phase
transition will be seen in the present approximation. Consider
variational functions $G$, for which $\partial^2G^2$
is short range, like in equation (\ref{solution}). The interaction between the
strings is then short range. In this case the standard energy vs. entropy
argument is telling us that at large $g^2$ the strings are condensed.
The string gas contribution to the vacuum energy will then be sizable.
Since this contribution has a negative sign, this situation
will be energetically
favored. Therefore at large $g^2$ one expects the best
variational state to have a short ranged $\partial^2G^2$, and therefore
the mass scale will be generated dynamically.
At small $g^2$ the vortex rings do not condense even for short range
interaction between them.
Their contribution to the vacuum energy will be negligible,
and one expects that the best variational vacuum will be determined by the
contributions from the $A_i$ and $\tilde \phi$ integrals, which will
lead to the same solution as in the noncompact theory.

{\bf Acknowledgements.}
A.K. thanks T. Bhattacharya, Y. Kluger,
E. Marino, M. Polikarpov and B.Rosenstein
for useful discussions. We are especially grateful
to Y. Kluger for participation in different stages of this work
and B. Rosenstein for making a crucial
remark at a critical moment.

\renewcommand{\footnotesize}{\small}

\noindent

\bigskip

{\renewcommand{\Large}{\normalsize}

\end{document}